# Phase-field approach to polycrystalline solidification including heterogeneous and homogeneous nucleation

Tamás Pusztai,† György Tegze,‡ Gyula I. Tóth,† László Környei,† Gurvinder Bansel,‡ Zhungyun Fan,‡ and László Gránásy‡§

† Research Institute for Solid State Physics and Optics, P O Box 49, H–1525 Budapest, Hungary
‡ Brunel Centre for Advanced Solidification Technology, Brunel University, Uxbridge, UB8 3PH, UK
§ Corresponding author: Laszlo.Granasy@brunel.ac.uk or grana@szfki.hu

**Abstract.** Advanced phase-field techniques have been applied to address various aspects of polycrystalline solidification including different modes of crystal nucleation. The height of the nucleation barrier has been determined by solving the appropriate Euler-Lagrange equations. The examples shown include the comparison of various models of homogeneous crystal nucleation with atomistic simulations for the single component hard-sphere fluid. Extending previous work for pure systems (Gránásy L, Pusztai T, Saylor D and Warren J A 2007 *Phys. Rev. Lett.* **98** art no 035703), heterogeneous nucleation in unary and binary systems is described via introducing boundary conditions that realize the desired contact angle. A quaternion representation of crystallographic orientation of the individual particles (outlined in Pusztai T, Bortel G and Gránásy L 2005 *Europhys. Lett.* **71** 131) has been applied for modeling a broad variety of polycrystalline structures including crystal sheaves, spherulites and those built of crystals with dendritic, cubic, rhombododecahedral, truncated octahedral growth morphologies. Finally, we present illustrative results for dendritic polycrystalline solidification obtained using an atomistic phase-field model.

## 1. Introduction

A substantial fraction of the technical materials used in the everyday life are polycrystalline, i.e., are composed of crystallites whose size, shape and composition distributions determine their macroscopic properties and failure characteristics of these substances (Cahn 2001). Size of the constituent crystallites may range from nanometers to centimeters in different classes of materials. While polycrystalline materials are the subject of an intensive research for some time, many aspects of polycrystalline solidification are still little understood. The complexity of multigrain crystallization is exemplified by thin polymer layers, which show an enormous richness of crystallization morphologies (Geil 1996). Polycrystalline morphologies of particular interest are the ubiquitous multi-grain dendritic and spherulitic structures. The multi-grain dendritic structures are composed of a large number of pine-tree-like dendritic crystals (relatives of the ice-flowers forming on window panes) and besides a broad range of other materials they have been seen in crystallizing colloidal suspensions (Cheng *et al* 2002). The term 'spherulite' is used in a broader sense for densely branched, polycrystalline solidification patterns (Magill 2001). Besides polymers and biopolymers, they have been seen in a broad variety of systems including alloys, mineral aggregates and volcanic rocks, liquid crystals, oxide and metallic glasses, even chocolate and biological systems. In particular, the world of minerals provides beautifully complex examples of such structures (Shelf and Hill 2003). Appearance of semi-crystalline spherulites of amyloid fibrils is associated with the Alzheimer and Creutzfeldt-Jakob diseases, type II. diabetes, and a range of systemic and neurotic disorders (Jin *et al* 2003, Krebs *et al* 2005), kidney stones of polycrystalline spherulitic structure have been observed (Khan *et al* 1979, Lambert *et al* 1998), and the formation kinetics of ice crystals influence the extent of damage biological tissues undergo during freezing (Zacharaissen and Hammel 1988). Other remarkably complex polycrystalline morphologies appear in composite materials, such as the "shish-kebab" structure in carbon nanotube containing polymers (Li *et al* 2006), and the plate like branched structures (e.g, graphite in cast iron and in other systems (Napolitano *et al* 2004, Hyde *et al* 2004)). Crystallization can be influenced by intrinsic and external fields such as composition, temperature, pressure, flow and electromagnetic fields. For example, modulated fields have been used to influence the dendritic crystallization morphology both in experiment and modeling: flow (Bouissou *et al* 1990), laser pulses (Quian and Cummins 1990, Murray *et al* 1995), and pressure (Börzsönyi *et al* 1999, 2000, Koss *et al* 2005, Li *et al* 2007). Although in the present paper, we concentrate mainly on techniques that are able to address polycrystalline solidification in specific intrinsic (composition) and external fields (temperature), we consider possible inclusion of other fields (e.g., flow).

First, we need, however, a suitable model of polycrystalline solidification that incorporates crystal nucleation and growth on equal footing. The fact that very similar polycrystalline morphologies are seen in substances of very different molecular geometry raises the hope that a coarse grained field theoretic model that neglects the molecular details might be able to capture some of the essential factors that govern crystalline pattern formation in such systems. It is expected that nucleation, diffusional instabilities, crystal symmetries, and the presence of particulate impurities play an important role. A particularly interesting mode of polycrystalline solidification, identified recently, is *growth front nucleation* (GFN), where growth takes place via continuous formation of new grains at the solidification front, a growth mechanism typical for spherulites and fractal-like polycrystals (Gránásy *et al* 2004a 2004b 2005). Accordingly, the model needs to address homogeneous and heterogeneous nucleation of growth centers and growth front nucleation (homogeneous and heterogeneous) together with diffusional instabilities.

Advances in computational materials science offer various methods to model polycrystalline solidification, which include cellular automata (e.g., Zhu and Hong 2002, Beltram-Sanchez and Stefanescu 2004, Zhu *et al* 2008), level set (e.g., Tryggvason *et al* 2001, Tan and Zabaras 2006 2007) and other front tracking techniques (e.g., Schmidt 1996, Steinbach *et al* 1999, Jacot and Rappaz 2002), and phase-field approaches (recent reviews: Boettinger *et al* 2002, Chen 2002, Hoyt *et al* 2003, Gránásy *et al* 2004a). Among them the phase-field models appear to be perhaps the most popular ones as they connect thermodynamic and kinetic properties with microstructure via a transparent mathematical formalism. In the phase-field theory, the local state of matter is characterized by a non-conserved structural order parameter $\phi(r, t)$, called phase field, which monitors the transition between the solid and liquid states. The time evolution of the structural order parameter is usually coupled to that of other slowly evolving conserved fields such as temperature or composition.

The phase-field model has already been used to determine the height of the nucleation barrier for homogeneous and heterogeneous nucleation (Gránásy *et al* 2002 2003a 2007) In the case of homogeneous nucleation a quantitative study has been performed for the hard sphere systems utilizing the thermodynamic, interfacial free energy and interface thickness data to fix the model parameters in equilibrium. Then the Euler-Lagrange equations have been solved to obtain the unstable equilibrium corresponding to crystal nuclei in the supersaturated state. This procedure delivers the free energy of nuclei *without adjustable parameters*, which can be then compared to the nucleation barrier data measured directly by atomistic simulations (Auer and Frenkel 2001a 2001b). It has been found that with an orientation averaged interfacial free energy of $\sim 0.61\ kT/\sigma^2$, obtained from molecular dynamics and Monte Carlo techniques (Davidchack and Laird 2000, Cacciuto *et al* 2003), a fair agreement can be seen with a phase-field model that relies on a quartic free energy and the usual intuitive but thermodynamically consistent interpolation function. Remarkably, recently, the free energy of the hard sphere crystal-liquid interface has been reduced considerably to $\sim 0.559\ kT/\sigma^2$ (Davidchack *et al* 2006) which certainly spoils the fair agreement (here $k$, $T$, and $\sigma$ are Boltzmann's constant, the temperature and the diameter of the hard spheres, respectively). It would be then natural to explore whether a better approximation could be obtained by using physically motivated double well and interpolation functions emerging from a Ginzburg-Landau expansion of the free energy (Gránásy and Pusztai 2002).

In the case of heterogeneous nucleation appropriate boundary conditions have been introduced at the foreign wall to realize the required contact angle (Gránásy *et al* 2007). Properties of the heterogeneous nuclei in two dimensions (2D) were obtained by solving numerically the respective Euler-Lagrange equation under these boundary conditions. This work needs too be extended to 3D and to alloys.

Modeling of polycrystalline solidification requires the inclusion of homogeneous and/or heterogeneous nucleation into the phase-field model. In field theoretic models it is done traditionally by adding Langevin noise of appropriate properties to the equations of motion (Gunton *et al* 1983). However, to describe the impingement of a large number of crystallites that grow *anisotropically*, one needs to incorporate the crystallographic orientations that allow the specification of the preferred growth directions of growth. The first phase-field model that introduces different crystallographic orientations into a solidifying system (Morin *et al* 1995) relies on a free energy density that has $n$ wells, corresponding to $n$ crystallographic orientations, breaking thus the rotational symmetry of the free energy. Simulations have been then performed to study polymorphous crystallization, where the composition of the liquid remains close to that of the crystal, therefore, chemical diffusion plays a minor role, and the system follows the Johnson-Mehl-Avrami-Kolmogorov (JMAK) kinetics (see e.g., Christian 1981). A weakness of the model is that the rotational invariance of the free energy density had to be sacrificed and a finite number of crystallographic orientations need to be introduced to enable the formation of grain boundaries of finite thickness.

A different approach for addressing the formation of particles with random crystallographic orientations is realized by the *multi-phase-field theory* (MPFT; see e.g., Steinbach *et al* 1996, Fan and Chen 1996, Tiaden *et al* 1998, Diepers *et al* 2002, Krill and Chen 2002), in which a separate phase field is introduced for every crystal grain. This models offer flexibility at the expense of enhanced mathematical/numerical complexity. MPFT has been used to study polycrystalline dendritic and eutectic/peritectic solidification, and has also been successfully applied for describing the time evolution of multigrain structures. However, the large number of phase fields applied in these approaches leads to difficulties, when nucleation is to be modeled by Langevin noise. While noise-induced nucleation can certainly be substituted by inserting nuclei by 'hand' into the simulations, this procedure becomes excessively non-trivial, when structures that require the nucleation of different crystallographic orientations at the growth front are to be addressed. Such a treatment, furthermore, rules out any possible interaction between diffusion and the orientation of new grains. In this way realization of growth front nucleation in the MPFT is not immediately straightforward.

It appears that modeling of complex polycrystalline structures and especially of GFN, requires another approach that relies on an orientation field to monitor the crystallographic orientation. The first model of this kind has been put forward by Kobayashi *et al* (1998) to model polycrystalline solidification in 2D, which uses a non-conserved scalar field to monitor crystallographic orientation. Assuming a free energy density of $f_{ori} = HT|\nabla\theta|$, where the coefficient $H$ has a minimum at the position of the interface, the minimization of free energy leads to a stepwise variation of $\theta(\mathbf{r})$, a behavior approximating reasonably the experimental reality of stable, flat grain boundaries. (Such minimum can be realized making the coefficient $H$ dependent on the phase field, e.g., by introducing the factor $1 - p(\phi)$ into $f_{ori}$ (Gránásy et al 2002)) Various modifications of this approach have been successfully applied for describing problems including solid-solid and solid-liquid interfaces (Kobayashi et al 1998, 2000, Warren et al 2003). A further important contribution was the modeling of the *nucleation of grains with different crystallographic orientations*, which has been solved by Gránásy et al (2002), who extended the orientation field $\theta$ into the liquid phase, where it has been made to fluctuate in time and space. Assigning local crystal orientation to liquid regions, even a fluctuating one, may seem artificial at first sight. However, due to geometrical and/or chemical constraints, a short-range order exists even in simple liquids, which is often similar to the one in the solid. Rotating the crystalline first-neighbor shell so that it aligns optimally with the local liquid structure, one may assign a local orientation to every atom in the liquid. The orientation obtained in this manner indeed fluctuates in time and space. The correlation of the atomic positions/angles shows how good this fit is. (In the model, the fluctuating orientation field and the phase field play these roles.) Approaching the solid from the liquid, the orientation becomes more definite (the amplitude of the orientational fluctuations decreases) and matches to that of the solid, while the correlation between the local liquid structure and the crystal structure improves. $f_{ori} = [1 - p(\phi)]|\nabla\theta|$ recovers this behavior by realizing a strong coupling between the orientation and phase fields. This addition to the orientation field model, first introduced by Gránásy et al (2002), facilitates the quenching of orientational defects into the crystal, leading to a mechanism generating new grains at the growth front. Indeed this approach of ours successfully describes the formation of such complex polycrystalline growth patterns formed by GFN as disordered ("dizzy") dendrites (Gránásy et al 2003b), spherulites (Gránásy et al 2003c 2004a 2004b 2005), 'quadrites' (Gránásy et al 2005), fractallike aggregates (2004a) and eutectic grains with preferred orientation between the two crystalline phases (Lewis et al 2004). The generalization of this approach to three dimensions has been done somewhat later. Practically at the same time two essentially equivalent formulations have been put forward: Pusztai et al (2005a 2005b) used the quaternion representation for the crystallographic orientation in solidification problems, while Kobayashi and Warren (2005a 2005b) proposed a rotation matrix representation to address grain boundary dynamics. A shortcoming of these earlier works is that crystal symmetries have not been taken into account in the simulations, although Pusztai et al (2005a 2005b) outlined in their papers how crystal symmetries should be handled in grain boundary formation.

A promising new field theoretic formulation of polycrystalline solidification is the *Phase-Field Crystal* (PFC) model (Elder et al 2002 2007, Elder and Grant 2004), which addresses freezing on the atomistic/molecular scale. The PFC approach is a close relative of the classical density functional theory (DFT) of crystallization: one may derive it by making a specific approximation for the two-particle direct correlation function of the liquid (Elder and Grant 2004, Elder et al 2007) in the Ramakrishnan-Yussouff expansion of the free energy functional of the crystal relative to the homogeneous liquid (for review on DFT see Oxtoby 1991). Remarkably, the PFC description includes automatically the elastic effects and crystal anisotropies, while addressing interfaces, dislocations and other lattice defects on the atomic scale. It has the advantage over traditional atomistic simulations (such as molecular dynamics), that it works on the diffusive time scale, i.e., processes taking place on about a million times longer time scale than molecular dynamics can address. The PFC method has already demonstrated its high potential for modeling dendrites, eutectic structures, polycrystalline solidification, grain boundaries / dislocations, epitaxial growth, crack formation, etc. (Elder and Grant 2004, Elder et al 2007, Provatas et al 2007). However, due to its atomistic nature it cannot be easily used to model large scale polycrystalline structures. Combination of a coarse grained formulation of the binary PFC theory based on the renormalization group technique outlined for the single component case with adaptive mesh techniques (Goldenfeld et al 2005, Athreya et al 2006, 2007) will certainly enhance the simulation domain for multi-component systems in the future. Another difficulty is that the crystal lattice and the respective anisotropy of the interfacial free energy cannot be easily tuned, although recent work incorporating three-body correlation opens up the way for advance in this direction (Tupper and Grant 2008). While the PFC is undoubtedly an excellent tool for investigating the atomistic aspects of polycrystalline solidification, it cannot easily address such scale morphologies as 3D multi-grain dendritic structures or spherulites: they seem to belong yet to the domain of conventional phase-field modeling. With appropriate numerical techniques, however, the PFC model might be applicable to address even such problems under specific conditions in 2D.

Herein, we apply the phase-field method to address various aspects of nucleation and polycrystalline solidification: (i) We reassess phase-field models of homogeneous crystal nucleation in the hard sphere system. (ii) We determine the structure and the barrier height for heterogeneous nucleation in a binary alloy. (iii) We apply the model of Pusztai et al (2005a 2005b) for describing polycrystalline solidification while considering crystal symmetries in handling the orientation field (crystallites with orientations related to each other by symmetry operations should not form grain boundaries), and demonstrate that the model is able to describe complex polycrystalline solidification morphologies based on dendritic, cubic, rhombododecahedral, and truncated octahedral growth forms, besides the transition between single needle crystals and polycrystalline spherulites. We combine the model with boundary conditions that realize pre-defined contact angles which is then used to model the formation of shish-kebab structures on nano-fibers. We introduce then a spatially homogeneous flow and a fixed temperature gradient to mimic directional solidification, which is then used to model the columnar to equiaxed transition in a binary alloy. (iv) Finally, we model multi-grain dendritic solidification in the framework of the binary PFC approach.

## 2. Phase-field models used

*2.1 Phase-field approach to nucleation barrier in homogeneous and heterogeneous nucleation*

As in other continuum models the critical fluctuation or nucleus represents an extremum of the appropriate free energy functional, therefore can be found by solving the respective sets of Euler-Lagrange equations. In the following we present phase-field models for two cases: (a) Homogeneous nucleation in the hard sphere system that crystallizes to the fcc (face centered cubic) structure, where besides the structural changes, we explicitly incorporate the density change during crystallization. (b) Heterogeneous nucleation in a binary system, where appropriate boundary conditions will be introduced to fix the contact angle in equilibrium.

*2.1.1 Phase-field model of homogeneous nucleation in the hard sphere system* Here we consider two possible phase-field approaches. Following previous work (Gránásy *et al* 2003a), the grand potential of the inhomogeneous system relative to the initial liquid is assumed to be a local functional of the phase field $m$ monitoring the liquid-solid transition ($m = 0$ and 1 in the liquid and in the solid, respectively) and the volume fraction $\varphi = (\pi/6)\sigma^3\rho$ (here $\rho$ is the number density of the hard spheres):

$$\Delta\Omega = \int d^3r \left\{ \frac{\varepsilon^2 T}{2}(\nabla m)^2 + \Delta\omega(m,\varphi) \right\}, \tag{1}$$

where, $\varepsilon$ is a coefficient that can be related to the interfacial free energy and the interface thickness, $T$ is the temperature, while $\Delta\omega(m,...)$ is the local grand free energy density relative to the initial state (that includes the Lagrange multiplier term, ensuring mass conservation; here the Lagrange multiplier is related to the chemical potential of the initial liquid). The gradient term leads to a diffuse crystal-liquid interface, a feature observed both in experiment (e.g, Howe 1996, Huisman *et al* 1997, Howe and Saka 2004, van der Veen and Reichert 2004) and computer simulations (e.g, Broughton and Gilmer 1986, Laird and Haymet 1992, Davidchack and Laird 1998, Ramalingam *et al* 2002). In the present work, grand potential density is assumed to have the following simple form:

$$\Delta\omega(m,\varphi) = wTg(m) + [1 - p(m)]f_S(\varphi) + p(m)f_L(\varphi) - \{\partial f_L/\partial\varphi\}(\varphi_\infty)[\varphi - \varphi_\infty] - f_L(\varphi_\infty), \tag{2}$$

where $f_S(\varphi)$ and $f_L(\varphi)$ are the Helmholtz free energy densities for the solid and liquid states, while $\varphi_\infty$ is the volume fractions of the initial (supersaturated) liquid phase. Different "double well" $g(m)$ and "interpolation" functions $p(m)$ will be used as specified below. The free energy scale $w$ determines the height of the free energy barrier between the bulk solid and liquid states. Once the functional forms of $g(m)$ and $p(m)$ are specified, model parameters $\varepsilon$ and $w$ can be expressed in terms of $\gamma_\infty$ and the thickness $\delta$ of the equilibrium planar interface (Cahn and Hilliard 1958).

Here we use two sets of these functions. One of them has been proposed intuitively in an early formulation of the PFT and in use widely:

*(a) The "standard" set (PFT/S):* These functions are assumed to have the form $g(\phi) = \frac{1}{4}\phi^2(1 - \phi)^2$ and $p(\phi) = \phi^3(10 - 15\phi + 6\phi^2)$, respectively, that emerge from an intuitive formulation of the PFT (Wang *et al* 1993). Here $\phi = 1 - m$ is the complementing phase field, defined so that it is 0 in the solid and 1 in the liquid. The respective expressions for the model parameters are as follows: $\varepsilon_S^2 = 6 \cdot 2^{1/2}\gamma_\infty\delta/T_f$, and $w_S = 6 \cdot 2^{1/2}\gamma_\infty/(\delta T_f)$. This model has been discussed in detail in (Gránásy *et al* 2003a).

*(b) Ginzburg-Landau form for fcc structure (PFT/GL):* Recently, we have derived these functions for bcc (base cetered cubic) and fcc (face centered cubic) structures (Gránásy and Pusztai 2002) on the basis of a single-order-parameter Ginzburg-Landau (GL) expansion that considers the crystal symmetries (Shih *et al* 1987). This treatment yields $g(m) = (1/6)(m^2 - 2m^4 + m^6)$ and $p(m) = 3m^4 - 2m^6$ for the fcc structure, while the expressions that relate the model parameters to measurable quantities are as follows: $\varepsilon_{GL}^2 = (8/3)C\varepsilon_S^2$, $w_{GL} = w_S(4C)^{-1}$, where $C = \ln(0.9/0.1) [3\ln(0.9/0.1) - \ln(1.9/1.1)]^{-1}$. Combination of the latter double well and interpolation functions with equation (2) is a new construction, presented here for the first time. Therefore, though it is analogous to the procedure applied in a previous work (Gránásy *et al* 2003a), we briefly outline the way the properties of nuclei are determined in this case:

The field distributions, that extremize the free energy, can be obtained solving the appropriate Euler-Lagrange (EL) equations:

$$\frac{\delta\Omega}{\delta m} = \frac{\partial I}{\partial m} - \nabla\frac{\partial I}{\partial\nabla m} = 0, \qquad \text{and} \qquad \frac{\delta\Omega}{\delta\varphi} = \frac{\partial I}{\partial\varphi} - \nabla\frac{\partial I}{\partial\nabla\varphi} = 0, \tag{3a,b}$$

where $\delta\Omega/\delta m$ and $\delta\Omega/\delta\varphi$ stands for the first functional derivative of the grand free energy with respect to the fields $m$ and $\varphi$, respectively. Here, $I = \frac{1}{2}\varepsilon^2T(\nabla m)^2 + f(m, \varphi) + \lambda\varphi$ is the total free energy density inclusive the term with a Lagrange multiplier $\lambda$ ensuring mass conservation, while the Helmholtz free energy density is $f(m, \varphi) = wTg(m) + [1 -$ $p(m)] f_S(\varphi) + p(m) f_L(\varphi)$. For the sake of simplicity, we assume here an isotropic interfacial free energy (a reasonable approximation for simple liquids). Note that due to a lack of a gradient term for the field $\varphi$ in the grand potential, equation (3b) yields an implicit relationship between $m$ and $\varphi$, which can be then inserted into equation (3a), when solving it.

Herein, equation (3b) has been solved numerically, using a variable fourth/fifth order Runge-Kutta method (Korn and Korn 1970), assuming an unperturbed liquid ($m = 0$, $\varphi = \varphi_\infty$) in the far field ($r \to \infty$), while, for symmetry reasons, a zero field gradient applies at the center of the fluctuations. Since $m$ and $dm/dr$ are fixed at different locations, the central value of $m$ that leads to $m \to m_\infty = 0$ for $r \to \infty$, have been determined iteratively. Having determined the solutions $m(r)$ and $\varphi(r)$, the work of formation of the nucleus, $W$ has been obtained by inserting these solutions into the grand potential difference (equation (1)).

Of these two phase-field models (PFT/S and PFT/GL), the latter, which relies on the Ginzburg-Landau expansion, incorporates a more detailed physical information on the system (e.g., crystal structure), therefore, it is expected to provide a better approximation to the atomistic simulations.

The physical properties, we use here, are the same as in a previous work of us (Gránásy et al 2003a), with the exception of the 10%-90% interface thickness, which is now allowed to change between $3.0\sigma$ and $3.3\sigma$, values that are consistent with the interfacial profiles for a variety of physical properties (such as coarse grained density, diffusion, and orientational order parameters $q_4$ and $q_6$) at the equilibrium solid-liquid interface of the hard sphere system (Davidchack and Laird 1998). In Section 3.1, we are going to address uncertainties associated with the interface thickness and interfacial free energy taken from atomistic simulations.

*2.1.2 Phase-field model of heterogeneous nucleation in binary alloys* Here, we have two fields to describe the local state of the matter, the usual phase field $\phi(\mathbf{r})$ and the concentration field $c(\mathbf{r})$. In the order to keep the problem mathematically simple, we assume again an isotropic solid-liquid interface. Then the Euler-Lagrange equation can be solved in a cylindrical coordinate system. Furthermore, if we do not assume a gradient term for the concentration field in the free energy, in equilibrium, there exists an explicit relationship between the phase field and the local concentration. Under these conditions, we need to solve the following Euler-Lagrange equation for the phase field

$$\frac{1}{2}\frac{\partial}{\partial r}\left(r\frac{\partial \phi}{\partial r}\right) + \frac{\partial^2 \phi}{\partial z^2} = \frac{p'(\phi)\Delta f[\phi,c] + g'(\phi)wT}{\varepsilon^2 T}, \qquad (4)$$

while in the absence of a $|\nabla c|^2$ term in the free energy, the Euler-Lagrange equation for the concentration field yields a $c(\phi)$ relationship. Accordingly, in equation (4) $\Delta f[\phi, c(\phi)] = f[\phi, c(\phi)] - (\partial f/\partial c)(c_\infty)[c(\phi) - c_\infty] - f_\infty$ is driving force of crystallization, while properties with subscript $\infty$ refer to quantities characterizing the initial liquid state. Now we wish to ensure in equilibrium (stable or unstable) that the solid-liquid interface has a fixed contact angle $\psi$ with a foreign wall placed at $z = 0$. To achieve this, we prescribe the following boundary condition at the wall, which can be viewed as a binary generalization of Model A presented in (Gránásy et al 2007):

$$(\mathbf{n} \cdot \nabla \phi) = \sqrt{\frac{2\Delta f[\phi, c(\phi)]}{\varepsilon^2 T}} \cos(\psi), \qquad (5)$$

where $\mathbf{n}$ is the normal vector of the wall. The motivation for this boundary condition is straightforward in the case of a stable triple junction, in which the equilibrium planar solid-liquid interface has a contact angle $\psi$ with the wall. The wall is assumed to lead to an ordering of the adjacent liquid, an effect that extends into a liquid layer of thickness $d$, which is only a few molecular diameters thick (see e.g., Toxvaerd 2002, Webb et al 2003). If we take plane $z = z_0$, which is slightly above this layer, i.e., $z_0 > d$, the structure of the equilibrium solid-liquid interface remains unperturbed by the wall (see figure 1). Then along the $z = z_0$ plane the phase field and concentration profiles are trivially related to the equilibrium profiles across the solid-liquid interface. Evidently, in the interface the following relationship holds

$$\frac{\varepsilon^2 T}{2}\left(\frac{\partial \phi}{\partial n_{SL}}\right)^2 = \Delta f[\phi, c(\phi)], \qquad (6)$$

where $n_{SL}$ is a spatial coordinate normal to the solid-liquid interface, while the component of $\nabla\phi$ normal to the wall is then $(\mathbf{n}\cdot\nabla\phi) = (\partial\phi/\partial n_{SL})\cdot\cos(\psi) = [2\Delta f/(\varepsilon^2 T)]^{1/2}\cdot\cos(\psi)$. (Remarkably, if in equilibrium a parabolic groove approximation by Folch and Plapp (2003 2005) is applied for the free energy surface, one finds that conveniently $\Delta f[\phi, c(\phi)] = wTg(\phi)$.) While equation (5) is straightforward for the equilibrium planar solid-liquid interface, generalization of this approach for nuclei involves further considerations. Indeed, in the undercooled state the planar interface is not in equilibrium, $\Delta f[\phi, c(\phi)]$ is a tilted double well, and equation (6) is not valid anymore. Note that it is the capillary pressure that restores the uniform chemical potential inside the nucleus (being in unstable equilibrium). While, in principle, it would be possible to solve the appropriate spherical Euler-Lagrange equation for the phase field, and use the respective solution to determine the normal component $P_N(\phi)$ of the pressure tensor that makes the chemical potential spatially uniform, it seems rather unpractical. It turns out, however, that at least for large nuclei (small undercoolings) a fairly good approximation can obtained if equation (5) is retained, however, with $\Delta f' = \Delta f - [1 - p(\phi)] \cdot \Delta f_0$, where $\Delta f_0$ is the driving force of solidification in the undercooled state. Note that the correction term mimics the effect of capillary pressure.

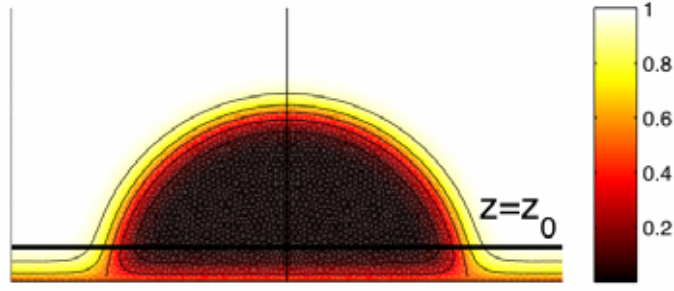

**Figure 1.** Typical cross-sectional phase-field map of a nucleus if the structural effects of a wall placed at $z = 0$ are considered (computation performed with Model B, Warren proposed in Gránásy *et al* (2007)). Note the boundary layers between the wall and the solid phase ($\phi = 0$), and between the wall and the liquid phase ($\phi = 1$). Note also that the crystal becomes disordered at the wall, while an ordering of the liquid takes place near the wall. Above plane $z = z_0$ the solid liquid interface remains unperturbed by the presence of the wall. In the case of a stable triple junction, however, the solid-liquid interface will be planar (not curved as for nuclei).

*2.2. Polycrystalline phase-field theory with quaternion representation of crystallographic orientations*

Here we use the three-dimensional PF model of polycrystalline solidification (Pusztai *et al* 2005a 2005b). Besides the usual square-gradient and local free energy density terms, the free energy functional consists of an orientational contribution:

$$F = \int d^3r \left\{ \frac{\varepsilon_\phi^2 T}{2} |\nabla \phi|^2 + f(\phi, c, T) + f_{ori} \right\}. \tag{7}$$

The local physical state of the matter (solid or liquid) is characterized by the phase field $\phi$ and the solute concentration $c$, while $\varepsilon_\phi$ is a constant, and $T$ is the temperature. The local free energy density is assumed to have the form $f(\phi, c, T) = w(\phi)T\,g(\phi) + [1 - p(\phi)]\,f_S(c) + p(\phi)\,f_L(c)$, where the intuitive "double well" and "interpolation" functions shown in the section 2.1.1 are used, while the free energy scale is $w(\phi) = (1 - c)\,w_A + c\,w_B$. The respective free energy surface has two minima ($\phi = 0$ and $\phi = 1$, corresponding to the crystalline and liquid phases, respectively), whose relative depth is the driving force for crystallization and is a function of both temperature and composition as specified by the free energy densities in the bulk solid and liquid, $f_{S,L}(c,T)$, taken here for the binary systems from the ideal solution model, or from CALPHAD type computations (computer aided CALculation of PHAse Diagrams).

The orientational contribution to free energy $f_{ori}$ has been obtained as follows. In 3D, the relative orientation with respect to the laboratory system is uniquely defined by a single rotation of angle $\eta$ around a specific axis, and can be expressed in terms of the three Euler angles. However, this representation has disadvantages: It has divergences at the poles $\vartheta = 0$ and $\pi$, and one has to use trigonometric functions that are time consuming in numerical calculations. Therefore, we opt for the four symmetric Euler parameters, $q_0 = \cos(\eta/2)$, $q_1 = c_1 \sin(\eta/2)$, $q_2 = c_2 \sin(\eta/2)$, and $q_3 = c_3 \sin(\eta/2)$, a representation free of such difficulties. (Here $c_i$ are the components of the unit vector $c$ of the rotation axis.) These four parameters $q = (q_0, q_1, q_2, q_3)$, often referred to as a *quaternion*, satisfy the relationship $\sum_i q_i^2 = 1$, therefore, can be viewed as a point on the four-dimensional (4D) unit sphere (Korn and Korn 1970). (Here $\sum_i$ stands for summation with respect to $i = 0, 1, 2,$ and 3, a notation used throughout this paper.)

The angular difference $\delta$ between two orientations represented by quaternions $q_1$ and $q_2$ can be expressed as $\cos(\delta) = \frac{1}{2}[\mathrm{Tr}(\mathbf{R}) - 1]$, where the matrix of rotation $\mathbf{R}$ is related to the individual rotation matrices $\mathbf{R}(q_1)$ and $\mathbf{R}(q_2)$, that rotate the reference system into the corresponding local orientations, as $\mathbf{R} = \mathbf{R}(q_1) \cdot \mathbf{R}(q_2)^{-1}$. After lengthy but straightforward algebraic manipulations one finds that the angular difference can be expressed in terms of the differences of quaternion coordinates: $\cos(\delta) = 1 - 2\Delta^2 + \Delta^4/2$, where $\Delta^2 = (q_2 - q_1)^2 = \sum_i \Delta q_i^2$, is the square of the Euclidian distance between the points $q_1$ and $q_2$ on the 4D unit sphere. Comparing this expression with the Taylor expansion of the function $\cos(\delta)$, one finds that $2\Delta$ is indeed an excellent approximation of $\delta$. Relying on this approximation, we express the orientational difference as $\delta \approx 2\Delta$.

The free energy of small-angle grain boundaries increases approximately linearly with the misorientation of the neighboring crystals, saturating at about twice of the free energy of the solid-liquid interface. Our goal is to reproduce this behavior of the small angle grain boundaries. To penalize spatial changes in the crystal orientation, in particular the presence of grain boundaries, we introduce an orientational contribution $f_{ori}$ to the integrand in equation (1), which is invariant to rotations of the whole system. While in 2D, the choice of the orientational free energy in the form $f_{ori} =$ $HT[1 - p(\phi)]|\nabla\theta|$ (where the grain boundary energy scales with $H$) ensures a narrow grain boundary and describes successfully both polycrystalline solidification and grain boundary dynamics (Kobayashi *et al* 1998 2000, Warren *et al* 2003, Gránásy *et al* 2002 2004a 2004b), in 3D we postulate an analogous intuitive form

$$f_{ori} = 2HT[1 - p(\phi)]\left\{\sum_i (\nabla q_i)^2\right\}^{1/2}. \tag{8}$$

It is straightforward to prove that this form boils down to the 2D model, provided that the orientational transition across grain boundaries has a fixed rotation axis (perpendicular to the 2D plane) as assumed in the 2D formulation.

As in 2D, to model crystal nucleation in the liquid, we extend the orientation fields, $q(\mathbf{r})$, into the liquid, where they are made to fluctuate in time and space. Note that $f_{ori}$ consists of the factor $[1 - p(\phi)]$ to avoid double counting of the orientational contribution in the liquid, which is *per definitionem* incorporated into the free energy of the bulk liquid. With appropriate choice of the model parameters, an ordered liquid layer surrounds the crystal as seen in atomistic simulations.

Time evolution of the field is assumed to follow relaxation dynamics described by the equations of motion:

$$\dot{\phi} = -M_\phi \frac{\delta F}{\delta \phi} + \zeta_\phi = M_\phi\left\{\nabla\left(\frac{\partial I}{\partial \nabla\phi}\right) - \frac{\partial I}{\partial \phi}\right\} + \zeta_\phi, \tag{9a}$$

$$\dot{c} = \nabla M_c \nabla\left(\frac{\delta F}{\delta c} - \zeta_j\right) = \nabla\left\{\frac{v_m}{RT}Dc(1-c)\nabla\left[\left(\frac{\partial I}{\partial c}\right) - \nabla\left(\frac{\partial I}{\partial \nabla c}\right) - \zeta_j\right]\right\}, \tag{9b}$$

$$\frac{\partial q_i}{\partial t} = -M_q \frac{\delta F}{\delta q_i} + \zeta_i = M_q\left\{\nabla\left(\frac{\partial I}{\partial \nabla q_i}\right) - \frac{\partial I}{\partial q_i}\right\} + \zeta_i. \tag{9c}$$

Here $I$ is the integrand of the free energy functional, $v_m$ is the molar volume, $D$ the diffusion coefficient in the liquid, and $\zeta_i$ are the appropriate noise terms representing the thermal fluctuations. (Conserved noise for the conserved fields and non-conserved noise for the non-conserved fields (Karma and Rappel 1999).) The time scales for the fields are determined by the mobility coefficients appearing in the coarse-grained equations of motion: $M_\phi$, $M_c$ and $M_q$. These coarse-grained mobilities can be taken from experiments and/or evaluated from atomistic simulations (see e.g., Hoyt *et al* 2003). For example, the mobility $M_c$, is directly proportional to the classic inter-diffusion coefficient for a binary mixture, the phase-field mobility $M_\phi$ dictates the rate of crystallization, while the orientational mobility $M_q$ controls the rate at which regions reorient, a parameter that can be related to the rotational diffusion coefficient and is assumed to be common for all quaternion components. While the derivation of a more detailed final form of equations (9a) and (9b) is straightforward, in the derivation of the equations of motion (equations 9(c)) for the four orientational fields $q_i(\mathbf{r})$, we need to take into account the quaternion properties ($\sum_i q_i^2 = 1$), which can be done by using the method of Lagrange multipliers, yielding

$$\frac{\partial q_i}{\partial t} = M_q\left\{\nabla\left(HT[1-p(\phi)]\frac{\nabla q_i}{[\sum_l(\nabla q_l)^2]^{1/2}}\right) - q_i\sum_k q_k\nabla\left(HT[1-p(\phi)]\frac{\nabla q_k}{[\sum_l(\nabla q_l)^2]^{1/2}}\right)\right\} + \zeta_i. \tag{10}$$

Gaussian white noises of amplitude $\zeta_i = \zeta_{S,i} + (\zeta_{L,i} - \zeta_{S,i})p(\phi)$ are then added to the orientation fields so that the quaternion properties of the $q_i$ fields are retained. ($\zeta_{L,i}$ and $\zeta_{S,i}$ are the amplitudes in the liquid and solid, respectively.)

This formulation of the model is valid for triclinic lattice without symmetries (space group P1). In the case of other crystals, the crystal symmetries yield equivalent orientations that do not form grain boundaries. In previous works, we have proposed that the crystal symmetries can be taken into account, when discretizing the differential operators used in the equations of motions for the quaternion fields. Calculating the angular difference between a central cell and its neighbors, all equivalent orientations of the neighbor have to be considered, the respective angular differences $\delta$ be calculated (using matrices of rotation $\mathbf{R'} = \mathbf{R} \cdot \mathbf{S}_j \cdot \mathbf{R}^{-1}$, where $\mathbf{S}_j$ is a symmetry operator), of which the smallest $\delta$ value shall be used in calculating the differential operator. (For cubic structure, there are 24 different $\mathbf{S}_j$ operators, if mirror symmetries whose interpretation in continuum models is not straightforward are omitted.)

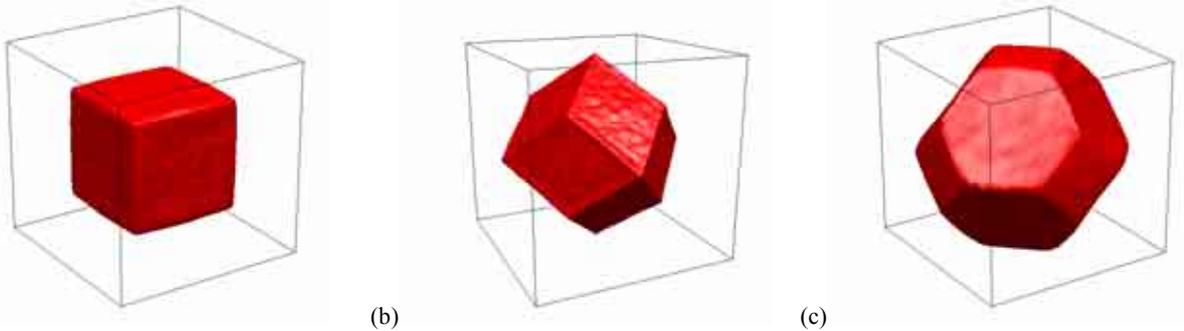

(a)          (b)          (c)

**Figure 2.** Single crystal growth forms at various choices of the anisotropy parameters of the kinetic coefficient: (a) cube ($\varepsilon_1 = -1.5$, $\varepsilon_2 = 0.3$); (b) rhombo-dodecahedron ($\varepsilon_1 = 0.0$, $\varepsilon_2 = 0.6$); (c) truncated octahedron ($\varepsilon_1 = 0.0$, $\varepsilon_2 = -0.3$). Here $\varepsilon_1$ and $\varepsilon_2$ are the coefficients of the fist and second terms in the Kubic harmonic expansion of the kinetic anisotropy.

Solving these equations numerically in three dimensions with an anisotropic interfacial free energy

$$\frac{\gamma(\mathbf{n})}{\gamma_0} = S(\mathbf{n}) = 1 + \varepsilon_1 \left( \sum_{i=1}^{3} n_i^4 - \frac{3}{5} \right) + \varepsilon_2 \left( \sum_{i=1}^{3} n_i^4 + 66 n_1^2 n_2^2 n_3^2 - \frac{17}{7} \right), \quad (11)$$

or with an anisotropic phase-field mobility of similar form $M_\phi = M_{\phi,0} S(\mathbf{n})$, one may obtain various single crystal growth forms as exemplified in figure 2. Note that in equation (11) $\mathbf{n} = (n_1, n_2, n_3)$ in the normal vector of the solid-liquid interface that can be expressed in terms of components of $\nabla \phi$.

*2.2.1 Boundary conditions* Unless stated otherwise, we have used periodic boundary condition in all directions. On foreign surfaces, a binary generalization of the boundary condition of Model A of (Gránásy et al 2007) has been applied (see also section 2.1.2). We model directional solidification by imposing a temperature gradient (note the excess term that appears because of the temperature dependent coefficient of $|\nabla \phi|^2$) and a uniform flow velocity in the simulation window. Foreign particles of given size and contact angle distributions of random lateral position and random crystallographic orientation were let in on the side, where high temperature liquid enters the simulation window.

*2.2.2 Materials properties* The polycrystalline calculations have been performed with three sets of materials parameters: (i) For an ideal solution approximant of the Ni-Cu system, we used in previous studies (for details see Pusztai et al 2005a). (ii) For a parabolic groove approximation of the free energy (developed by Folch and Plapp 2003 2005) adopted to the Ni-Cu system at 1574 K by Warren (2007). (iii) For the Al-Ti alloy of thermodynamic properties from a CALPHAD type assessment of the phase diagram (for details see Pusztai et al 2006).

*2.2.3 Numerical solutions* The equations of motion have been solved numerically using an explicit finite difference scheme. Periodic boundary conditions were applied. The time and spatial steps were chosen to ensure stability of our solutions. The noise has been discretized as described by Karma and Rappel (1999). A parallel codes relying on the MPI/OpenMPI protocols have been developed.

*2.3. Binary phase-field crystal model*

In derivation of the binary PFC, the starting point is the free energy functional of the binary perturbative density functional theory, where the free energy is Taylor expanded relative to the liquid state (denoted by subscript L) up to 2nd order in density difference (up to two-particle correlations) (Elder et al 2007):

$$\frac{F}{kT} = \int d\mathbf{r} \left[ \rho_A \ln\left(\frac{\rho_A}{\rho_A^L}\right) - \Delta\rho_A + \rho_B \ln\left(\frac{\rho_B}{\rho_B^L}\right) - \Delta\rho_B \right]$$
$$- \frac{1}{2} \iint d\mathbf{r}_1 d\mathbf{r}_2 [\Delta\rho_A(\mathbf{r}_1) C_{AA}(\mathbf{r}_1, \mathbf{r}_2) \Delta\rho_A(\mathbf{r}_2) + \Delta\rho_B(\mathbf{r}_1) C_{BB}(\mathbf{r}_1, \mathbf{r}_2) \Delta\rho_B(\mathbf{r}_2) + 2\Delta\rho_A(\mathbf{r}_1) C_{AB}(\mathbf{r}_1, \mathbf{r}_2) \Delta\rho_B(\mathbf{r}_2)] \quad (12)$$

where $k$ is Boltzmann's constant, $\Delta\rho_A = \rho_A - \rho_A^L$ and $\Delta\rho_B = \rho_B - \rho_B^L$. It is assumed here that all two point correlation functions are isotropic, i.e., $C_{ij}(\mathbf{r}_1, \mathbf{r}_2) = C_{ij}(|\mathbf{r}_1 - \mathbf{r}_2|)$. Taylor expanding direct correlation functions in Fourier space up to 4th order, one obtains $C_{ij} = [C_{ij}^0 - C_{ij}^2 \nabla^2 + C_{ij}^4 \nabla^4] \delta(\mathbf{r}_1 - \mathbf{r}_2)$ in real space, where $\nabla$ differentiates with respect to $\mathbf{r}_2$ (see Elder et al 2007). The partial direct correlation functions $C_{ij}$ can be related to measured or computed partial structure factors (see e.g. Woodhead-Galloway and Gaskell 1968).

Following Elder et al (2007), we introduce the reduced partial number densities differences $n_A = (\rho_A - \rho_A^L)/\rho_L$ and $n_A = (\rho_B - \rho_B^L)/\rho_L$, where $\rho_L = \rho_A^L + \rho_B^L$. It is also convenient to introduce the new variables $n = n_A + n_B$ and $(\delta N) = (n_B - n_A) + (\rho_B^L - \rho_A^L)/\rho_L$. Then, expanding the free energy around $(\delta N) = 0$ and $n = 0$ one obtains

$$\frac{F}{\rho_L kT} = \int d\mathbf{r} \left\{ \frac{n}{2} [B_L + B_S (2R^2 \nabla^2 + R^4 \nabla^4)] n + \frac{t}{3} n^3 + \frac{v}{4} n^4 + \gamma(\delta N) + \frac{w}{2} (\delta N)^2 + \frac{u}{4} (\delta N)^4 + \frac{L^2}{2} |\nabla(\delta N)|^2 + ... \right\}. \quad (13)$$

Assuming substitutional diffusion between species A and B, i.e., the same $M$ mobility applies for the two species, the dynamics of $n$ and $(\delta N)$ fields decouple. Assuming, furthermore, that the mobility is a constant $M_e$, the respective equations of motions have the form (Elder et al 2007):

$$\frac{\partial n}{\partial t} = M_e \nabla^2 \frac{\delta F}{\delta n} \quad \text{and} \quad \frac{\partial (\delta N)}{\partial t} = M_e \nabla^2 \frac{\delta F}{\delta (\delta N)}, \tag{14}$$

where $\frac{\delta F}{\delta \chi} = \frac{\partial I}{\partial \chi} + \sum_j (-1)^j \nabla^j \frac{\partial I}{\partial \nabla^j \chi}$ is the first functional derivative of the free energy with respect to field $\chi$, and $I$ is the integrand of equation (13), while the respective effective mobility is $M_e = 2M/\rho^2$. Expanding $B_L$, $B_S$ and $R$ in terms of ($\delta N$) with coefficients denoted as $B_j^L$, $B_j^S$ and $R_j$, assuming that only coefficient $B_0^L$, $B_2^L$, $B_0^S$, $R_0$ and $R_1$ differ from zero, and inserting the respective form of $I$ into equations (14), one finds

$$\frac{\partial n}{\partial t} = M_e \nabla^2 \left[ \begin{array}{l} n\{B_0^L + B_2^L (\delta N)^2\} + tn^2 + vn^3 + \frac{B_0^S}{2}\{2[R_0 + R_1(\delta N)]^2 \nabla^2 + [R_0 + R_1(\delta N)]^4 \nabla^4\}n \\ + \frac{B_0^S}{2}\{2\nabla^2(n[R_0 + R_1(\delta N)]^2) + \nabla^4(n[R_0 + R_1(\delta N)]^4)\} \end{array} \right], \tag{15a}$$

$$\frac{\partial (\delta N)}{\partial t} = M_e \nabla^2 \left[ \begin{array}{l} B_2^L (\delta N) n^2 + 2 B_0^S n \{[R_0 + R_1(\delta N)] R_1 \nabla^2 + [R_0 + R_1(\delta N)]^3 R_1 \nabla^4\}n \\ + \gamma + w(\delta N) + u(\delta N)^3 - L^2 \nabla^2 (\delta N) \end{array} \right]. \tag{15b}$$

These equations have been solved numerically using a semi-implicit spectral method based on operator splitting (Tegze *et al* 2008) under periodic boundary condition on all sides after adding a conservative noise (a random flux) to them that represent the thermal fluctuations with an ultraviolet cut off at the inter-atomic spacing.

*2.4. Computational resources*

The parallel codes developed for the phase-field end phase-field crystal models have been run on three recently built PC clusters: two at the Research Institute for Solid State Physics and Optics, Budapest, Hungary, consisting of 160 and 192 CPU cores (80 dual core Athlon processors with 1 Gbit/s (normal Ethernet) communication, and 24 × 2 × 4 CPU core Intel processors equipped with 10 Gbit/s fast communication (Infiniband)), respectively, and third PC cluster at the Brunel Centre for Advanced Solidification Technology, Brunel University, West London, UK, consisting of 160 CPU cores (20 × 2 × 4 CPU core Intel processors) and 1 Gbit/s (normal Ethernet) communication.

## 3. Results and discussion

*3.1. Quantitative test of phase-field models of homogeneous crystal nucleation in the hard sphere system*

The predicted nucleation barrier heights are presented for the usual intuitive and the Ginzburg-Landau expanded double well and interpolation functions in figure 3 as a function of volume fraction. It has been found that the barrier heights predicted by the PFT with physically motivated free energy (PFT/GL) gives a considerably closer agreement with direct results from atomistic simulations (Auer and Frenkel 2001a 2001b) than the PFT model with a free energy surface relying on the usual intuitively chosen double well and interpolation function (PFT/S). It is also remarkable that the droplet model of the classical nucleation theory fails spectacularly. We note here that in a previous study (Gránásy *et al.* 2003a), we used an interface thickness determined by the envelope of the density peaks. We believe that the present choice of $\delta_{10\%-90\%} \in [3.0\sigma, 3.3\sigma]$, which has been deduced from profiles for several physical properties should be more reliable. It is worth noting also that the interfacial data from atomistic simulations might somewhat underestimate both the interfacial free energy and the interface thickness due to the limited size of such simulations, which leads to a long wavelength cut off in the spectrum of surface fluctuations. On the other hand, interfaces relevant to nucleation are of a size scale that is comparable to the size scale of atomistic simulations, so one might expect here only minor error from this source.

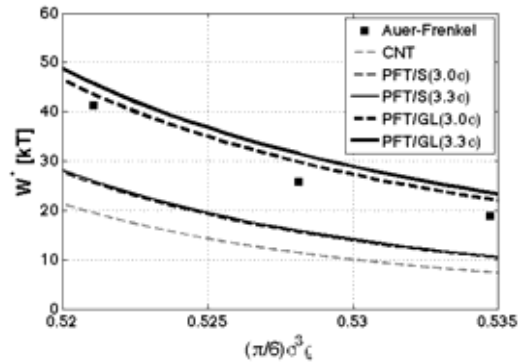

**Figure 3.** Comparison of the reduced nucleation barrier height ($W^*/kT$) vs volume fraction relationships various phase-field models predict for the hard sphere system without adjustable parameters. Predictions of PFT models with the intuitive (PFT/S) and with Ginzburg-Landau expanded (PFT/GL) double-well and interpolation functions are presented. There are two curves for each PFT model: one with the minimum (upper curve) and another with the maximum of the 10%–90% interface thickness deduced from atomistic simulations (Davidchack and Laird 1998). For comparison, direct results for $W^*$ from the Monte Carlo simulations (full squares; Auer and Frenkel 2001a 2001b), and parameter free predictions from the droplet model of the classical nucleation theory (CNT) are also shown.

*3.2. Structure and barrier for heterogeneous crystal nuclei in binary alloys*

The structure of the heterogeneous nuclei forming at 1574 K in a NiCu liquid alloy (with a free energy surface approximated by a parabolic groove (Foch and Plapp 2003)) of composition $(c - c_S)/(c_L - c_S) = 0.2$ under nominal contact angles $\psi = 30°, 60°, 90°, 120°,$ and $170°$ at a horizontal wall enforced by the boundary condition equation (5) are shown in figure 4. Note that the interface thickness is considerably smaller than the radius of curvature. Accordingly, in the non-wetting limit ($\psi \to \pi$), the height of the nucleation barrier can be approximated well with that from the classical droplet model of homogeneous nuclei. However, towards ideal wetting the nuclei are made almost entirely of interface, so the classical spherical cap model is expected to break down. Despite this, an analysis of the contour lines corresponding to $\phi = \frac{1}{2}$ gives contact angles within about 2° of the nominal (scattering with roughly this value). It is thus demonstrated that so far as the height of the nucleus is larger than the interface thickness the true contact angle falls reasonably close to the nominal value, i.e., the boundary condition given by equation (5) can be used with confidence to simulate surfaces of pre-defined contact angle of $\psi$.

It is also of interest to compare the nucleation barriers from the phase-field theory and from the classical spherical cap model relying on a sharp interface (the homogeneous nucleus can also be obtained as doubling the barrier height for 90° contact angle). It appears that under the investigated conditions the catalytic potency factor $f(\psi) = W_{hetero}/W_{homo}$ follows closely the function $f(\psi) = (1/4)[2 - 3\cos(\psi) + \cos(\psi)^3]$ from the classical spherical cap model (see rightmost panel in figure 4). This is reasonable, since these nuclei, as mentioned above, are fairly classical since their radius of curvature is large compared to the interface thickness.

Next, we apply this technique in phase-field *simulations* of heterogeneous nucleation. First, we apply it for the solidification of a single component system (only equation (9a) is solved here). Noise induced heterogeneous nucleation has been simulated on complex surfaces of $\psi = 60°$ including stairs, a checkerboard modulated surface, rectangular grooves and randomly positioned spheres with random radius, while using the properties of pure Ni (figure 5). Also we incorporate results for a non-wetting brush ($\psi = 175°$) protruding from a wetting surface ($\psi = 60°$), while at the center of the simulated area a wetting stage ($\psi = 60°$) is placed that helps crystal nucleation (figure 6). A complex behavior is seen: if the brush is dense, no nucleation is possible on the horizontal surfaces only at the central stage, and after nucleation the crystal "crawls" on the tips of the non-wetting brush. If the distance between the fibers in the non-wetting brush increases crystal can climb down to the horizontal wetting surface, while if this distance between the non-wetting fibers is large enough, nucleation may take place on the horizontal surface. Simulations of this kind might find application in nano-patterning studies.

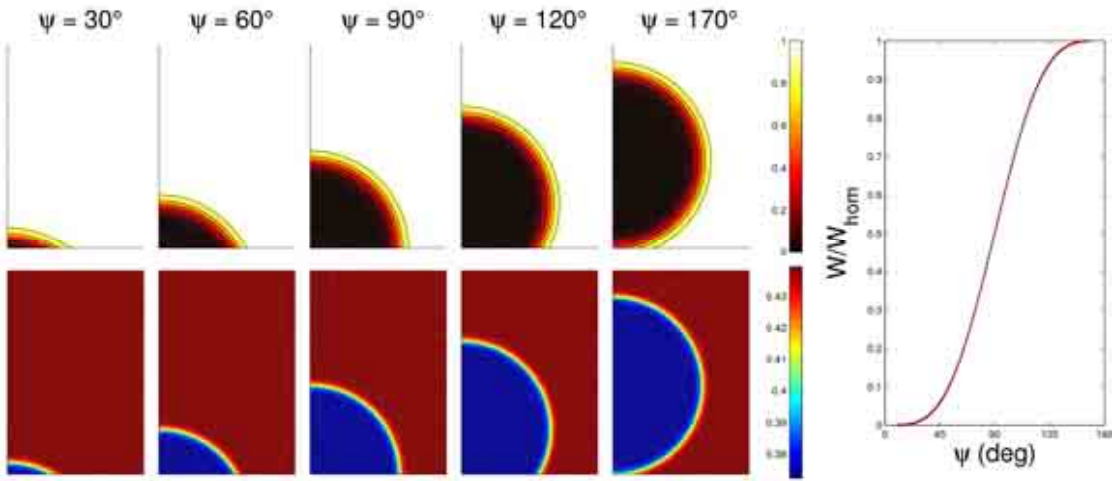

**Figure 4.** Phase-field (upper row) and composition (lower row) maps for heterogeneous nuclei obtained by solving numerically the respective Euler-Lagrange equation (equation (4)) as a function of contact angle $\psi$ in the binary NiCu system at 1574 K. The size of the calculation window is 100 × 150 nm. The contour lines in the upper row indicate phase field levels of $\phi = 0.1, 0.3, 0.5, 0.7,$ and 0.9, while the black contour line in the composition maps indicates the equilibrium composition of the solid phase $c_S^e = 0.399112$. Here parabolic well parameters corresponding to an interface thickness of 1.76 nm and a solid-liquid interfacial free energy of 0.3623 J/m$^2$ have been used. The classical (black) and non-classical (red) catalytic potency factors are shown on the right.

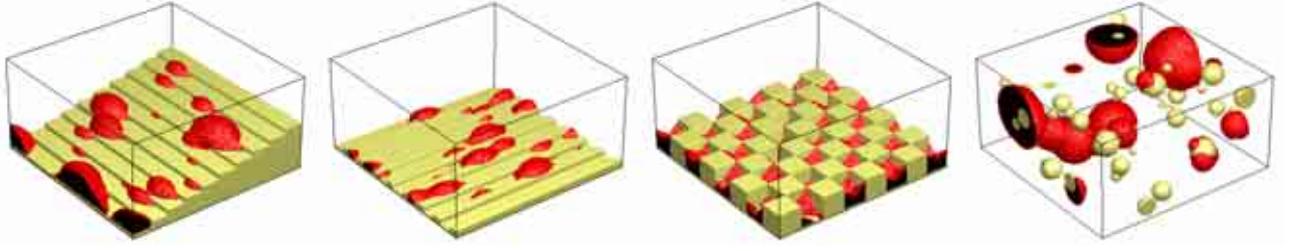

**Figure 5.** Noise induced heterogeneous crystal nucleation on complex surfaces of contact angle of 60°. From left to right: stairs, rectangular grooves, checkerboard modulated surface, and spherical particles. (Properties of Ni have been used.)

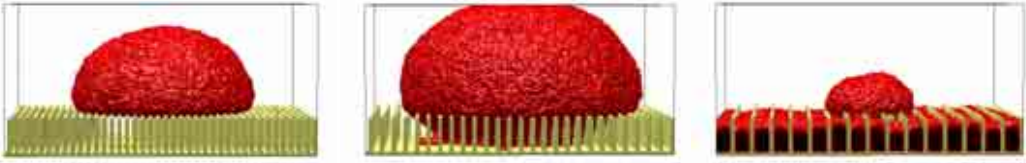

**Figure 6.** Crystal nucleation and growth on a non-wetting nano-fiber brush. Note the effect of decreasing density of the brush (from left to right) on crystallization. (For details see the text.)

*3.3. Modeling complex polycrystalline morphologies in three dimensions*

3D phase-field simulations showing the nucleation and growth of crystallites of different habits (cube, rhombo-dodecahedron, truncated octahedron, and dendritic) realized by prescribing appropriate kinetic anisotropies) illustrate the application of quaternion field for describing crystallographic orientation in figures 7 and 8. The physical properties of the Cu-Ni system has been used, the calculations were performed at 1574 K and at a supersaturation of $S = (c_L - c)/(c_L - c_S) = 0.75$, where $c_L = 0.466219$, $c_S = 0.399112$ and $c$ are the concentrations at the liquidus, solidus, and the initial homogeneous liquid mixture, respectively. The diffusion coefficient in the liquid was assumed to be $D_L = 10^{-9}$ m$^2$/s. Dimensionless mobilities of $M_{\phi,0} = 3.55 \times 10^{-1}$ m$^3$/Js (with an anisotropy of $M_\phi = M_{\phi,0}\{1 - 3\varepsilon_0 + 4\varepsilon_0 \; [(A_1\nabla\phi)_x^4 + (A_y\nabla\phi)_y^4 + (A_z\nabla\phi)_z^4]/|\mathbf{A}\nabla\phi|^4\})$ and $M_{q,L} = 8.17$ m$^3$/Js, and $M_{q,S} = 0$ were applied, while $D_S = 0$ was taken in the solid. The kinetics of multi-grain dendritic solidification has been simulated in a cube of 8.4 μm × 8.4 μm × 8.4 μm for a time interval of ~ 0.16 ms. The evolution of the normalized crystalline fraction $X$ has been analyzed in terms of the Johnson-Mehl-Avrami-Kolmogorov kinetics (Christian 1981), $X = 1 - \exp\{-(t/\tau)^{p_{AK}}\}$, where $\tau$ is a constant related to the nucleation and growth rates, and $p_{AK}$ is the Avrami-Kolmogorov exponent characteristic to the mechanism of transformation. The kinetic exponent evaluated from our simulations, $p_{AK} = 2.922 \pm 0.001$ ($\tau = 5 \times 10^{-5}$ s), falls between those for nucleation with diffusion controlled ($p_{AK} = 2.5$) and with steady state growth ($p_{AK} = 4$) (see Christian 1981). This implies that some of the particles have not yet reached the fully grown steady state dendritic morphology, as is apparent in figure 8. Larger simulations are planned to clarify further the relationship between morphology and $p_{AK}$. Note that here we have a reasonable statistics for nucleation, as by the end of the simulation, about 180 dendritic particles formed, a number considered sufficient for such purposes (Pusztai and Gránásy 1998).

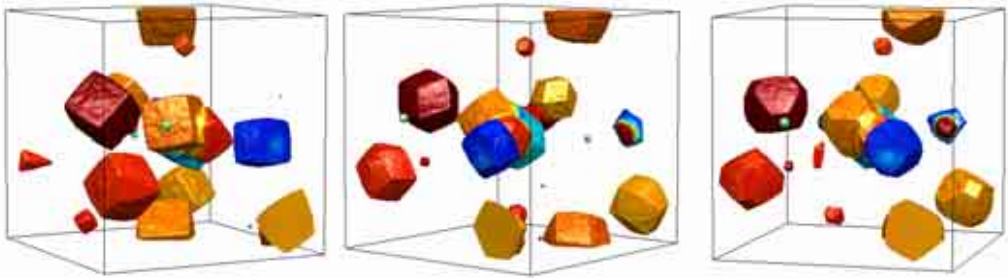

**Figure 7.** Polycrystalline structures formed by nucleation and growth of cubic, rhombo-dodecahedral and truncated octahedral crystals (from left to right, respectively). The computations have been performed on a 400 × 400 × 400 grid for ideal solution NiCu thermodynamics at 1574 K and supersaturation $S = 0.8$, however, with kinetic anisotropies given in the caption of figure 2.

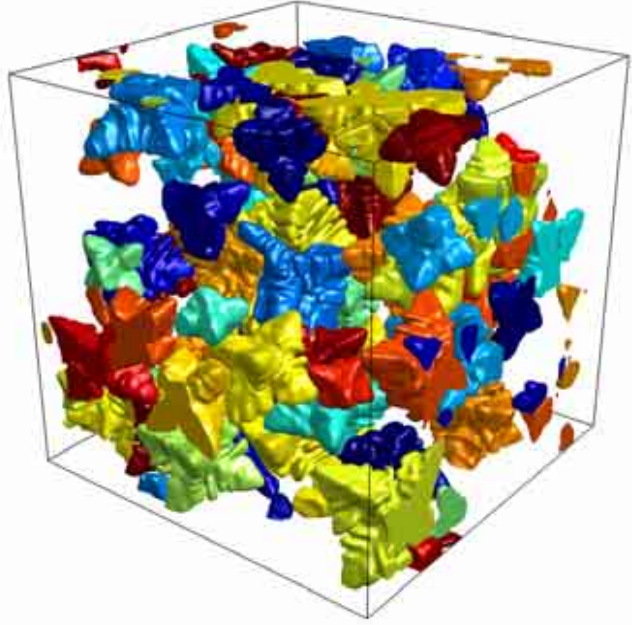

**Figure 8.** Polycrystalline structure formed by nucleation and dendritic growth in a NiCu alloy (whose thermodynamic properties were approximated by the ideal solution model) at 1574 K and $S = 0.78$, while assuming cubic crystal symmetries. Equations (9) have been solved numerically on a $640 \times 640 \times 640$ grid (~262 million grid points) by solving numerically the equations of motion (6 stochastic partial differential equations). The computation took about a month on 80 processors. By the end of the simulation, about 180 crystalline particles formed. Different colors indicate different crystallographic orientations.

As discussed in detail in previous work (Gránásy *et al* 2003c 2004a 2004b 2005), reduction of the orientation mobility in the case of needle crystals may lead to the formation of Category 2 spherulites, that start to grow as a single needle crystal but later the ends splay out and form eventually a space filling roughly spherical polycrystalline structure. A similar transition can be seen when increasing the driving force of solidification. As demonstrated in figure 9, the fre-quency by which new grains form at the ends of the needle crystal increases strongly with increasing supersaturation. The mechanism, by which the new grains form, is via quenching orientational defects into the solid, which defects might be identified as bunches of dislocations, as in 2D simulations (Gránásy *et al* 2006).

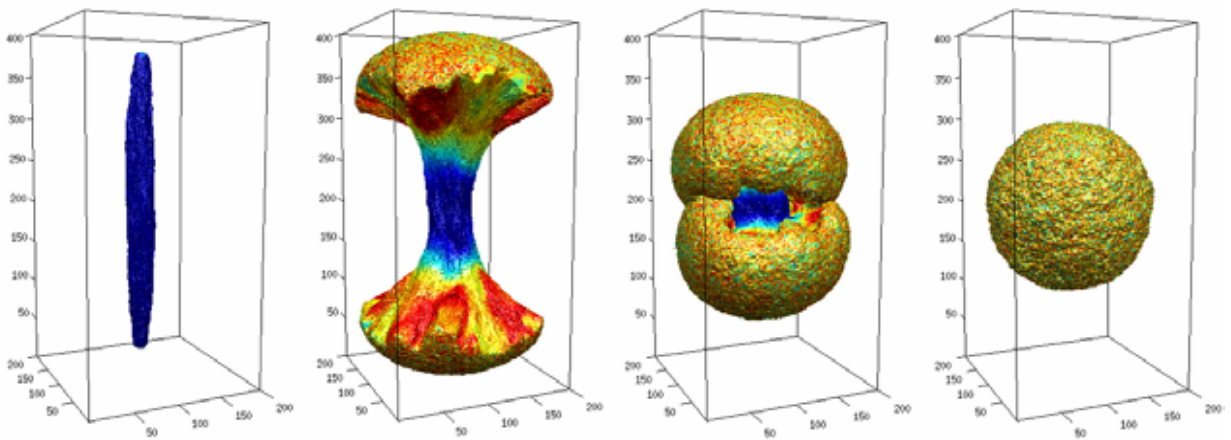

**Figure 9.** From needle crystal to spherulites in a phase-field theory relying on a quaternion representation of the crystallographic orientation. The simulations have been performed by solving equations (9) on a $200 \times 200 \times 400$ grid assuming ideal solution thermodynamics. A large kinetic anisotropy favoring a needle crystal form, characterized by the parameter values $\varepsilon_0 = 1/3$ and $\mathbf{A} = (0, 0, 1)$ has been applied. The driving force of solidification increases from the left to right ($S = 1.8, 1.9, 2.0,$ and $2.1$, respectively).

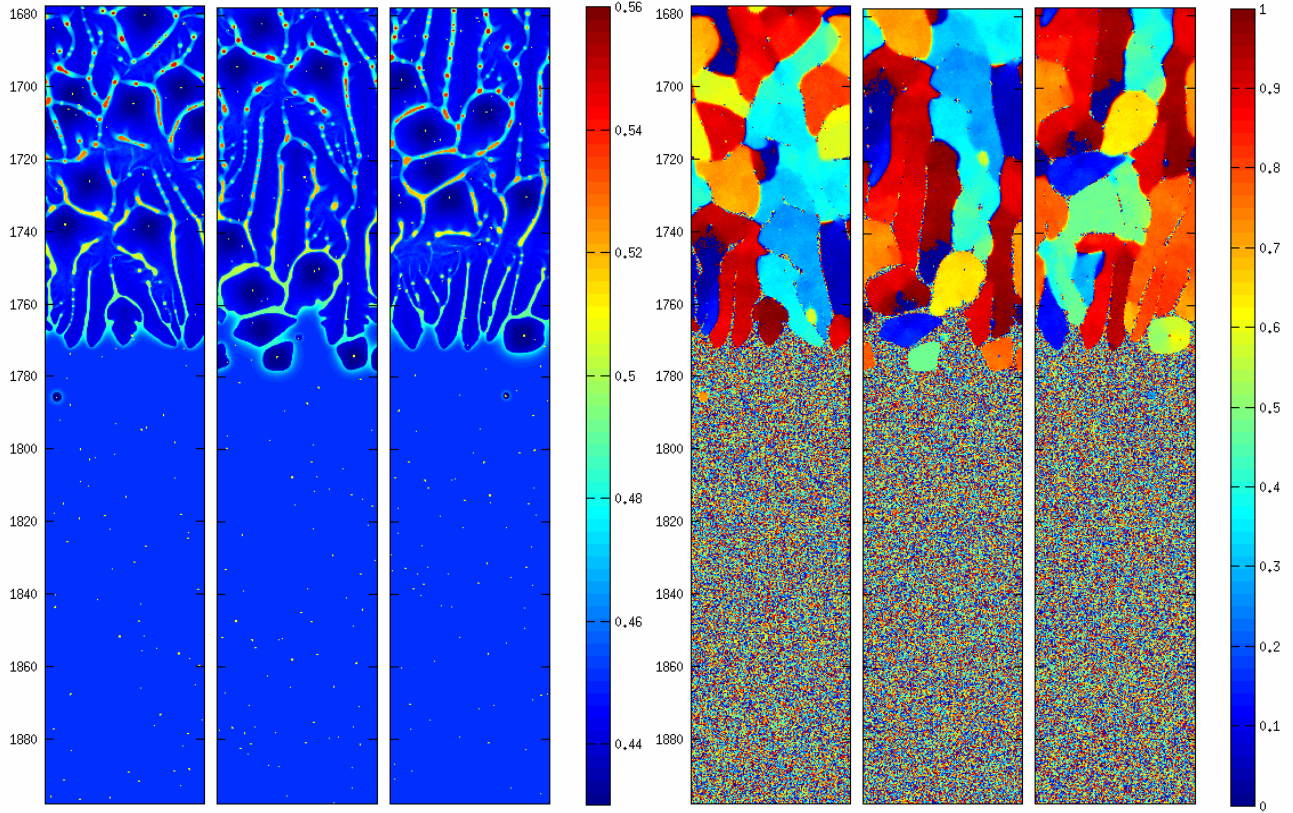

**Figure 10.** Phase-field simulation of polycrystalline solidification of the $Al_{0.45}Ti_{0.55}$ alloy in a moving frame ($V$ = 1.26 cm/s) and a constant temperature gradient ($\nabla T$ = 1.12 × $10^7$ K/m). Composition (on the left) and orientation maps (on the right) corresponding to times $t$ = 2.3, 2.6 and 2.9 ms are shown. Note that the orientations corresponding to 0 and 1 are equivalent. The computation has been performed by solving the 3D model (equations (9)) in 2D on a 600 × 3000 grid (3.93 μm × 19.69 μm). White spots in the chemical maps indicate the foreign particles, whose diameter varies in the 13 nm – 66 nm range, and have a contact angle of $\psi$ = 60°.

*3.3.1 Modeling of directional solidification* In order to model columnar to equiaxed transition (CET) in the framework of the EU FP6 IMPRESS project (Jarvis and Voss 2005), we have extended our 3D model to describe polycrystalline solidification of the $Al_{0.44}Ti_{0.55}$ alloy in temperature gradient and a moving frame. To enable large scale simulations, we have used a broad interface (65.6 nm), however, included an anti-trapping current (Kim *et al* 1999, Karma 2001, Kim 2007) to ensure a quantitative description of dendrites. In the simulation window, the material is made to move with a homogeneous velocity from the bottom to the top, while a fixed temperature gradient is prescribed in the vertical direction. Particles of given number density, random orientation and size, and of given contact angle are let to enter into the simulation window at the bottom edge. Snapshots of the chemical and orientation maps illustrating polycrystalline solidification under such conditions are presented in figure 10. As a result of the interplay between heterogeneous nucleation and growth, after the initial transient, we observe stochastically alternating nucleation-controlled and growth-controlled periods. This is a non-steady solution appearing in the CET zone. A detailed analysis of this phenomenon will be presented elsewhere (Pusztai *et al* 2008).

*3.3.2 Phase separation and polycrystalline solidification in the presence of fluid flow* In order to address the solidification of Al-Bi monotectic alloys (candidates for a new generation of self lubricating bearing materials), the 2D version of our polycrystalline phase-field theory has been combined with viscous flow. Solidification has been then modeled via introducing a phase-field dependent viscosity, and a non-classical stress-tensor related to the phase-, composition-, and orientation fields (Tegze and Gránásy 2006), while the regular solution model has been used to approximate the thermodynamics of the Al-Bi system. Inside the liquid-liquid immiscibility region we observed various hydrodynamic effects (Tegze *et al* 2005). Besides the solutal- and thermocapillary motion, we have seen flow assisted coagulation and bicontinuous phase separation (figure 11), mechanisms identified by Tanaka and coworkers (Tanaka 1995 1996, Tanaka and Araki 1998). It has also been found that the solute pile up ahead of the solidification front might significantly accelerate droplet nucleation in the metastable region of the liquid-liquid coexistence region (figure 11).

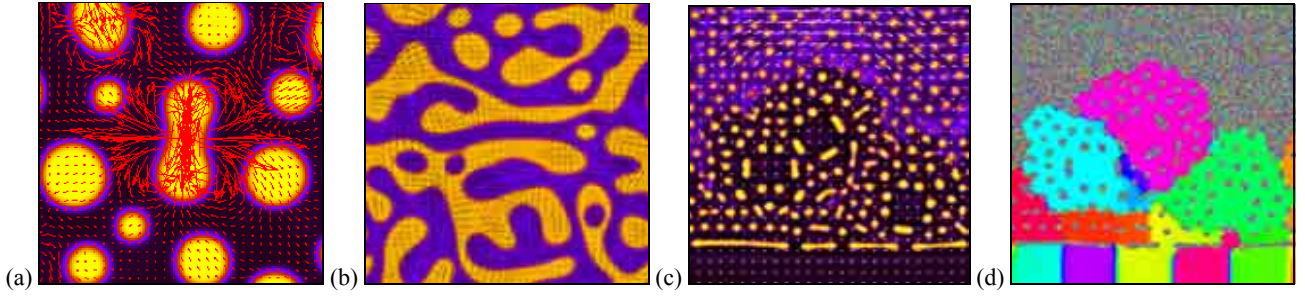

**Figure 11.** Liquid phase separation and solidification in monotectic alloys (regular solution approximant of Al-Bi): (a) Collision assisted collision of liquid droplets ($c_{Bi} = 0.25$, $T = 920$ K, $250 \times 250$ section of a $512 \times 512$ grid); (b) bicontinuous phase separation ($c_{Bi} = 0.5$, $T = 900$ K, $512 \times 512$ grid); (c),(d) solidification of phase separating liquid ($c_{Bi} = 0.23$, $T = 750$ K, $512 \times 512$ section of a $1024 \times 1024$ grid). Composition (a) – (c) and orientation maps (d) are shown. In panels (a) – (c) arrows indicate the velocity field.

*3.4. Atomistic simulations for polycrystalline solidification of a binary alloy in two dimensions*

We have performed simulations for the PFC model on a $16{,}384 \times 16{,}384$ grid using the same model parameters as Elder *et al* (2007), however, with half of their spatial step. Accordingly, our simulation window contains roughly 1.6 million atoms. Solidification has been initiated by inserting 5, 50 and 500 randomly oriented and positioned crystalline clusters of 13 atoms each into the simulation window. The resulting multi-grain structures are shown in figure 12 (snapshots of the "composition field" ($\delta N$) are displayed). The respective time dependencies of the number of atoms in the crystalline phase are presented in figure 13(a). The latter quantity has been obtained by counting the atoms in the crystalline state (an atom has been supposed to belong to the solid phase if its density peak was larger than the average of the value for the bulk liquid and the maximum value for the bulk crystal) using the public domain software ImageJ (Abramoff *et al* 2004). The higher level of crystalline fraction observed in the 500 particle simulation (~1.5 million atoms of a total of ~1.6 million) signals a more efficient solute trapping, probably attributable to fact that here the initial transient of fast growth rate represents a larger fraction of the total solidification time than for 50 or 5 particles. This is also consistent with the observation that the contrast of the ($\delta N$) field grows with time. The time evolution of crystallization has been analyzed in terms of the JMAK kinetics. The respective Avrami plots and the kinetic exponent vs reduced transformed fraction curves are displayed in figures 13(b) and (c). The Avrami plots are not linear, and the respective Avrami-Kolmogorov exponents ($p_{AK}$) vary with the transformed fraction (or time). Apart from an initial transient, the observed $p_{AK}$ values fall between the limiting values $p_{AK} = d/2 = 1$ and $p_{AK} = d = 2$ corresponding to diffusion controlled (conserved dynamics) and interface controlled growth of fixed number of particles in 2D (Christian 1981). Possible origin of the observed time dependencies of $p_{AK}$ is that due to mass conservation, and the differences in the densities of the crystal and liquid, the driving force for crystallization decreases as crystallization proceeds. Screening effects characteristic to highly anisotropic growth (Shepilov 1990, Shepilov and Baik 1994, Birnie and Weinberg 1995, Pusztai and Gránásy 1998) are also expected to influence transformation kinetics of the dendritic particles. Finally, we note that the behavior of the $p_{AK}(X)$ curve for the 5 dendritic particles reflects the small number of these particles, which cannot provide a satisfactory statistics for an accurate evaluation of the kinetic exponent. Unfortunately, significantly larger simulations for a large number of fully developed dendrites cannot be easily made with the present numerical technique and the hardware we used.

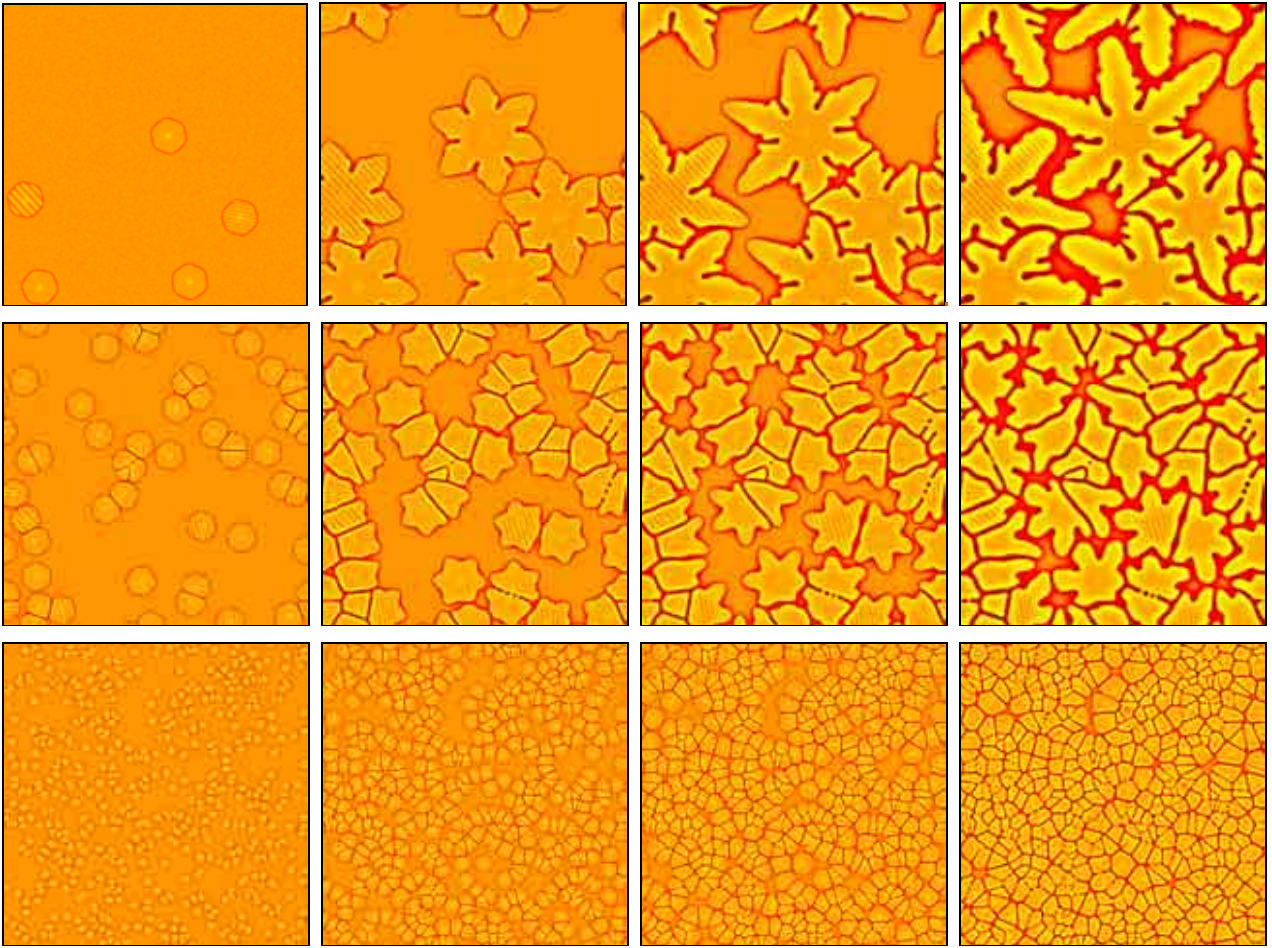

**Figure 12.** Polycrystalline solidification in the binary phase-field crystal model (the distribution of the ($\delta N$) field is shown). 1st row: Dendritic growth of 5 crystalline particles (snapshots taken at 1,000, 5,000, 10,000 and 20,000 time steps are shown). 2nd row: growth of 50 particles (snapshots taken at 1,000, 3,000, 5,000 and 10,000 time steps are shown). 3rd row: growth of 500 particles (snapshots taken at 250, 500, 750 and 1,500 time steps are shown). The simulations have been performed on a $16{,}384 \times 16{,}384$ grid, using a semi-implicit spectral method. Note that here the position of all atoms of the crystalline phase are known accurately.

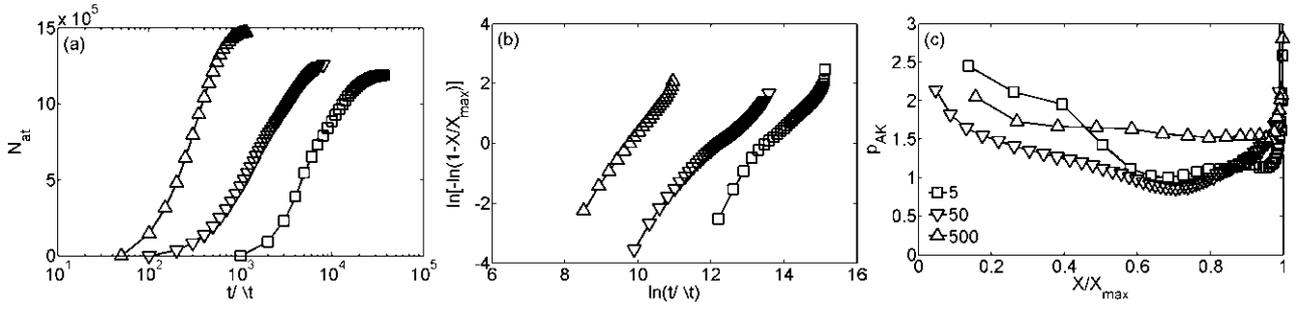

**Figure 13.** Crystallization kinetics for binary phase-field crystal simulations shown in figure 12. (a) Number of atoms in the crystalline phase vs number of time steps; (b) Avrami plots ($X$ and $X_{max}$ are the transformed fraction and its maximum; the slope of the curve is the Avrami-Kolmogorov exponent $p_{AK}$); and (c) the kinetic (Avrami-Kolmogorov) exponent as a function of the reduced transformed fraction. (Upward and downward pointing triangles and squares correspond to 500, 50 and 5 particles, respectively.)

## 4. Summary

Using various phase-field techniques, we have addressed diverse aspects of polycrystalline solidification, including homogeneous and heterogeneous nucleation of growth centers, and polycrystalline growth. Along these lines, we have shown that using a physically motivated (Ginzburg-Landau expanded) free energy in the phase-field approach, a reasonably accurate prediction can be obtained for the nucleation barrier of homogeneous crystal nucleation in the hard sphere system. We have then presented a method for incorporating walls of pre-defined contact angle into phase-field simulations, and demonstrated that rather complicated problems (heterogeneous nucleation on patterned surfaces/nanofiber brush) can be treated this way. Next, we have shown that phase-field models based on a quaternion representation of the crystallographic orientation are able to address the formation of fairly complex three dimensional polycrystalline structures, including multi-grain dendritic solidification and the formation of polycrystalline spherulites. The effect of temperature and flow fields on polycrystalline solidification has also been explored. Finally, we have used a recently developed atomistic approach, the "phase-field crystal" model, to investigate multi-grain dendritic crystallization in a binary liquid alloy. We believe that these modeling tools and their descendants/combinations supported by atomistic simulations and *ab initio* computations will find application in various branches of materials science and technology.

## Acknowledgments

L. G. thanks J. A. Warren and J. F. Douglas for discussions on heterogeneous nucleation and polycrystalline solidification, M. Plapp for illuminating discussions on noise, and K. R. Elder for communications regarding the PFC method. This work has been supported by contracts OTKA-K-62588, ESA PECS Nos. 98021, 98043 and 98059, and by the EU FP6 Project IMPRESS under Contract No. NMP3-CT-2004-500635. P. T. is a grantee of the Bolyai János Scholarship of the Hungarian Academy of Sciences. G. B. has been supported by an EPSRC grant.